\documentclass{sig-alternate}

\usepackage{times}
\usepackage{fancyhdr}
\usepackage{rotating}

\usepackage{graphicx}
\usepackage{amssymb}
\usepackage{amsmath}
\usepackage{url}
\usepackage{multicol}
\usepackage{epsf,psfrag}
\usepackage{indentfirst}
\usepackage{float}
\usepackage[lined,algonl,boxed]{algorithm2e}

\makeatletter
\newcommand\figcaption{\def\@captype{figure}\caption}
\newcommand\tabcaption{\def\@captype{table}\caption}
\makeatother

\usepackage{color}

\newcommand{\spacing}[2]{
   \renewcommand{\baselinestretch}{#2}
   \small\normalsize #1
      \setlength{\parskip}{0.6\baselineskip}
   \settowidth{\parindent}{xxxx}
   \setlength{\parindent}{#2\parindent}
   \setlength{\leftmargini}{\parindent}
   \setlength{\leftmarginii}{\parindent}
   \setlength{\leftmarginiii}{\parindent}
   \setlength{\footnotesep}{#2\footnotesep}
}

\begin{document}
\spacing{\normalsize}{1}

\title{Characterizing Video Responses in Social Networks}

\author{
   \begin{tabular}[t]{ccc}
   \sc\footnotesize{Fabricio Benevenuto$^\ddagger$}&
   \sc\footnotesize{Fernando Duarte$^\ddagger$}&
   \sc\footnotesize{Tiago Rodrigues$^\ddagger$}\\[-1mm]
      \tt\footnotesize{fabricio@dcc.ufmg.br}&\tt\footnotesize{fernando@dcc.ufmg.br}&\tt\footnotesize{tiagorm@dcc.ufmg.br}\\[2 mm]
      \end{tabular}\vspace{0.06cm}\\
      \begin{tabular}[t]{ccc}
   \sc\footnotesize{Virgilio Almeida$^\ddagger$}&
   \sc\footnotesize{Jussara Almeida$^\ddagger$}&
   \sc\footnotesize{Keith Ross$^\dagger$}\\[-1mm]
   \tt\footnotesize{virgilio@dcc.ufmg.br}&\tt\footnotesize{jussara@dcc.ufmg.br}&\tt\footnotesize{ross@poly.edu}\\
      \end{tabular}\vspace{0.25cm}\\
      {\small$^\ddagger$Computer Science Department, Federal University of Minas Gerais, Brazil}\vspace{0.03cm}\\
         {\small$^\dagger$Polytechnic University, Brooklyn, NY, USA}\vspace{0.1cm}\\
}

\maketitle
\begin{abstract}
Video sharing sites, such as YouTube, use video responses to enhance the
social interactions among their users. The video response feature allows
users to interact and converse through video, by creating a video sequence
that begins with an opening video and followed by video responses from other
users. Our characterization is over 3.4 million videos and 400,000 video
responses collected from YouTube during a 7-day period. We first analyze the
characteristics of the video responses, such as popularity, duration, and
geography. We then examine the social networks that emerge from the video
response interactions.
\end{abstract}




\section{Introduction}
\label{sec:intro}


Online social video networking sites allow for video-based communication among their users.  Video-based functions are offered as alternative to text-based
ones, such as video reviews for products, video ads and video responses~\cite{Shannon}. A video response feature allows users to interact and converse through
video, by creating a video sequence that begins with an opening video followed by video responses from other users.

In this paper, we present a thorough characterization of video response in a social networking site, namely YouTube.  Streaming over 3.4 billion videos a month
(as of December 2007) according to ComScore, YouTube is the most popular social video-based media network today generating high-volumes of Internet traffic.
Our characterization of over 3.4 million videos spanning a one week period is done at two different levels.  At the bottom level, video responses are
characterized in terms of their popularity, duration, geographical origin, and other features. At the top level, social networks created by the interactions
among users and videos are analyzed.  Such a characterization is of interest for three reasons.  The first is a technical reason, stemming from the necessity to
understand video-based communication, in order to evaluate new protocols and design choices for video services.  The second reason is the volume of traffic
generated by video responses. The total number of views of the videos analyzed in our data set exceeds 20 billion and 14\% of this total refer to views of
video responses.  The third reason is sociological, relating to social networking issues that influence the behavior of users interacting primarily with stream
objects, instead of textual content traditionally available on the Web.

\subsection{Contributions}
Based on an extensive sample collected from YouTube, a dataset that consists of over 3.4 million videos collected  over a one-week period, we provide a
statistical analysis of  video responses and a network analysis of how users make use of this functionality in social networking environment. The principal
contributions of our study can be summarized as follows:

\noindent $\bullet$ The distributions of video responses across responsive and responded users and  responded videos follow power laws. This means that the
majority of the video-based interactions in  YouTube occur between a small fraction of users.
Moreover, the majority of the video responses are concentrated on a small fraction of the videos.

\noindent $\bullet$ The durations of responded videos and video responses follow Weibull distributions, although the duration of responses are more skewed
towards shorter values. Moreover, there is a strong correlation between the duration of the responded video and the average duration of its responses.

\noindent $\bullet$ A significant fraction (27\%) of responses to a video are actually uploaded to YouTube before the original video is uploaded. However, about
42\% of all responses are posted within one month after the original video, and 17\% are posted at least 100 days after it. Thus, for some videos, new user
interactions are initiated longer after it was expected.

\noindent $\bullet$ The vast majority (99\%) of the responded videos triggered interactions for which each user participated only once, resembling what occurs
in a guest-book. Nevertheless, a few videos triggered very lively interactions, with each responsive user participating at least 3 times. 

\noindent $\bullet$ 40\% of the responded videos have a percentage of responses from the same country superior to 60\%, which may be exploited in the design of
content distribution mechanisms. 

\noindent $\bullet$ 35\% of the responded videos receive at least one self-response, that is, a response posted by the user who posted the original video.
However, the correlation between the number of video responses and the number of views of a responded video is weak. 

\noindent $\bullet$ The community structure of the social network formed by video responses possesses a small strongly connected component (SCC) and large
number of small communities (up to 20 members) and  singletons (one degree nodes).  We  found that the network created by users of a SCC is tightly connected,
with an average clustering coefficient equal to 0.137, which is much greater than 0.04 of the whole network.  Our findings suggest that the community structure
developed by video responses is at an early stage, for the size of the SCC is around 5\% of the network size.

\subsection{Related Work}

Given the novelty of video response, it is natural to ask whether its characteristics are similar to traditional videos and other objects of the web. Indeed,
over the last few years, there has been a number of studies that explored the various aspects of social networking sites. For example, the works
in~\cite{moon2007,mislove2007,arlitt2007} explored the overall scope, structure, and relation patterns of the popular online social networks: Flickr, YouTube,
LiveJournal, and Orkut. A study of YouTube is presented in~\cite{moon2007}.  Reference~\cite{facebook} studies temporal
access and social patterns in Facebook.  The authors analyze content characteristics and system issues that can be used to improve video distribution
mechanisms, such as caching and peer-to-peer distribution schemes.  Such studies are important because they allow us to understand characteristics of different
types of online social networks. Another important consideration relates to the patterns of accesses targeting social networks  and in particular how such
access patterns impact the portion of web traffic induced by social networking sites.  Unlike these studies, our analysis of YouTube focuses on characterizing
video responses in online social networks.  To the best of our knowledge, this is the first effort towards  understanding the video response functionality  in
social network.

\subsection{Paper Outline}
The rest of the paper is organized as follows. Next section  describes how we crawled and sampled YouTube. Section 3
presents a statistical characterization of video responses. In the following section we present a characterization 
of video interactions in the social network formed by relationships between users. 
We conclude and present directions for future research in Section~\ref{sec:conc}.

\vspace{-0.1cm}
\section{Crawling a Social Network}

To collect data, we visit pages on the YouTube site (that is, crawl) and gather information about YouTube video responses and their contributors.  Every YouTube
video post has a single contributor, who is a registered YouTube user.  We say a YouTube video is a {\em responded video} if it has at least one video response.
A responded video has a sequence of video responses listed chronologically in terms of when they were created. We say a YouTube user is a {\em responded user}
if at least one of its contributed videos is a responded video. Finally, we say that a YouTube user is a {\em responsive user} if it has posted at least one
video response.

A natural user graph emerges from video responses. At a given instant of time $t$, let $X$ be the union of all responded users and responsive users.  The set
$X$ is, of course, a subset of all YouTube users.  We denote the {\em video response user graph} as the directed graph $(X, Y)$, where $ (x_1,x_2)$ is a
directed arc in $Y$ if user $x_1 \in X$ has responded to a video contributed by user $x_2 \in X$.  In general, the video response user graph 
have multiple weakly connected components.  Some components may have thousands of users, and some components may have only two users.  Ideally, we would like to
obtain the complete video response user graph $(X,Y)$.  Although YouTube provides lists of the 100 most responded videos of all time, it does not currently
provide a means to systematically visit all the responded videos.  In particular, it is difficult to find the users in the small components in the graph
$(X,Y)$.

\incmargin{3em}
\restylealgo{boxed}\linesnumbered
\begin{algorithm}[t]
\scriptsize
   \SetLine
   \SetKwInOut{Input}{input}
   \SetKwInOut{Output}{output}
   \caption{Crawler}

   \BlankLine
   \Input{A list L of users (seeds)}
   \BlankLine
   \ForEach{User $U$ in L}{
      Collect $U$'s info using the YouTube API\;
      Collect $U$'s video list using the API\;
      \ForEach{Video $V$ in the video list}{
         Copy the HTML of $V$\;
         \If{$V$ is a responded video}{
            Copy the HTML of $V$'s video responses\;
            Insert the responsive users in L\;
         }
         \If{$V$ is a video response}{
            Insert the responded user in L\;
         }
      }
   }
   \BlankLine
   \label{alg:Crawler Strategy}
\end{algorithm}
\decmargin{3em}

Instead, we design a sampling procedure that allows us to obtain a large representative subset $A$ of $X$.  For a given sampled subset $A$ of $X$, let $(A,B)$ be
the directed graph, where $(a_1,a_2)$ is a directed arc in $B$ if user $a_1 \in A$ has responded to a video contributed by user $a_2 \in A$.  Note that sampled
graph $(A,B)$ is a sub-graph of $(X,Y)$. It is desirable that the sampled set of users $A$ has the following properties: 1) Each connected component in $(A,B)$
is a connected component in $(X,Y)$; that is, the sampled subgraph $(A,B)$ consists of (entire) connected components from $(X,Y)$. This property is important in
order to analyze the social networking interactions engendered by video responses.  2) The subset $A$ covers a large fraction of $X$ (at least 60\%). Our sample
would then include the majority of the responded and responsive users.  3) The most responded users are included in $A$. This property ensures that we are
including the most important users, and only neglecting users who have few responded videos.
To this end, we designed the sampling procedure described in Algorithm 1.

Starting with any seed set, the above algorithm ensures that our resulting sample graph $(A,B)$ has the
first property listed above, namely, $(A,B)$ consists of (some of the) connected components from the
entire video response user graph $(X,Y)$. We discuss the other properties subsequently.

We ran this sampling procedure with two different seed sets.  Our first seed set uses the contributors of the all-time top-100 responded videos. Our second
seed set is based on the random sampling technique described in Algorithm 2.

\incmargin{3em}
\restylealgo{boxed}\linesnumbered
\begin{algorithm}[b]
\scriptsize
   \SetLine
   \SetKwInOut{Input}{input}
   \SetKwInOut{Output}{output}
   \caption{Find random seeds}
   \BlankLine
   \Input{A list of words from a dictionary}

   \BlankLine
   Select a random word from the dictionary\;
   Search tag using YouTube API, using the word as tag\;
    \ForEach{Contributor $C$ of the videos found}{
       \If{$C$ is a responded \textbf{OR} a responsive user}{
           Add user to list\;
       }
   }
   \BlankLine
   \emph{Randomly select 100 users from the final list}\;
   \BlankLine
   \label{alg:random}
\end{algorithm}
\decmargin{3em}

Let $(A,B)$ be the sample graph determined with seed set consisting of the all-time top-100 most responded videos.
We now verify that our sampling procedures have Properties 2 and 3 in addition to
Property 1. Of the 100 random seed users, we find that 67 of those users belong to $A$.
Thus, our sampling scheme satisfies Property 2.
To verify Property 3, we use the second data set (generated with the 100 random seeds), and rank
order the 10 most, 100 most and 1000 most responded users. We find that our sampled graph
$(A,B)$ contains all 10 of the 10 most responded users, 98 of the 100 most responded users, and
951 of 1000 most responded users. Thus, Property 3 is verified as well.

\begin{table}[t]
\begin{center}
\scriptsize
\begin{tabular}{|r|r|r|}
\hline
\bf characteristic & \bf top-100 & \bf random \\
Period of sampling & 09/21-09/26/07 & 10/21-10/24/07 \\
\hline
\multicolumn{3}{|c|}{\bf video-data}\\
\hline
 \# videos collected & 3,436,139 & 3,974,579 \\
 \# video responses  & 417,759   & 485,716 \\
 \# views & 20,645,583,524 & 24,380,755,065 \\
 \# views of responses & 2,826,822,374 & 3,309,004,711 \\
 \# videos without response &  3,225,560 & 3,729,522 \\
\hline
Common videos & \multicolumn{2}{|c|}{3,112,660}\\
Common responded videos & \multicolumn{2}{|c|}{200,113}\\
Common video responses & \multicolumn{2}{|c|}{372,566}\\
\hline
\multicolumn{3}{|c|}{\bf User-data}\\
\hline
 \# users collected & 160,765 & 182,725 \\
\hline
Common users & \multicolumn{2}{|c|}{146,799}\\
\hline
\end{tabular}
\end{center}
\vspace{-0.2cm}
\caption {Summary of Crawled Data Sets}
\vspace{-0.2cm}
\label{tab:workload1}
\end{table}

The basic statistics of our two crawl are provided in Table 1. To summarize, our all-time top-100 video crawler collected data about 3.4 million videos and 160
thousands users.  We now discuss the data in more detail, focusing first on the video crawl, and then on the user crawl.

\begin{figure*}[t]
\vspace{-0.2cm}
\centering
\includegraphics[width=.27\textwidth]{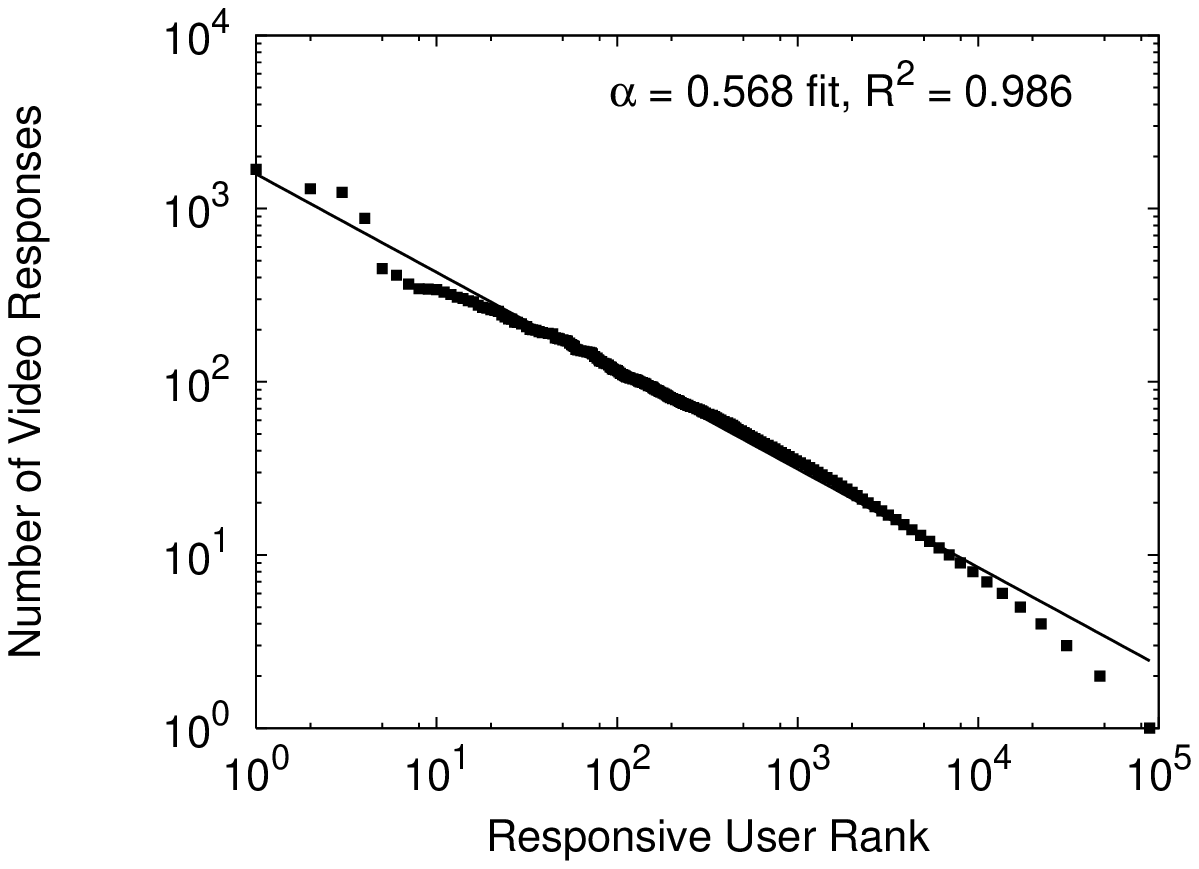}
\includegraphics[width=.27\textwidth]{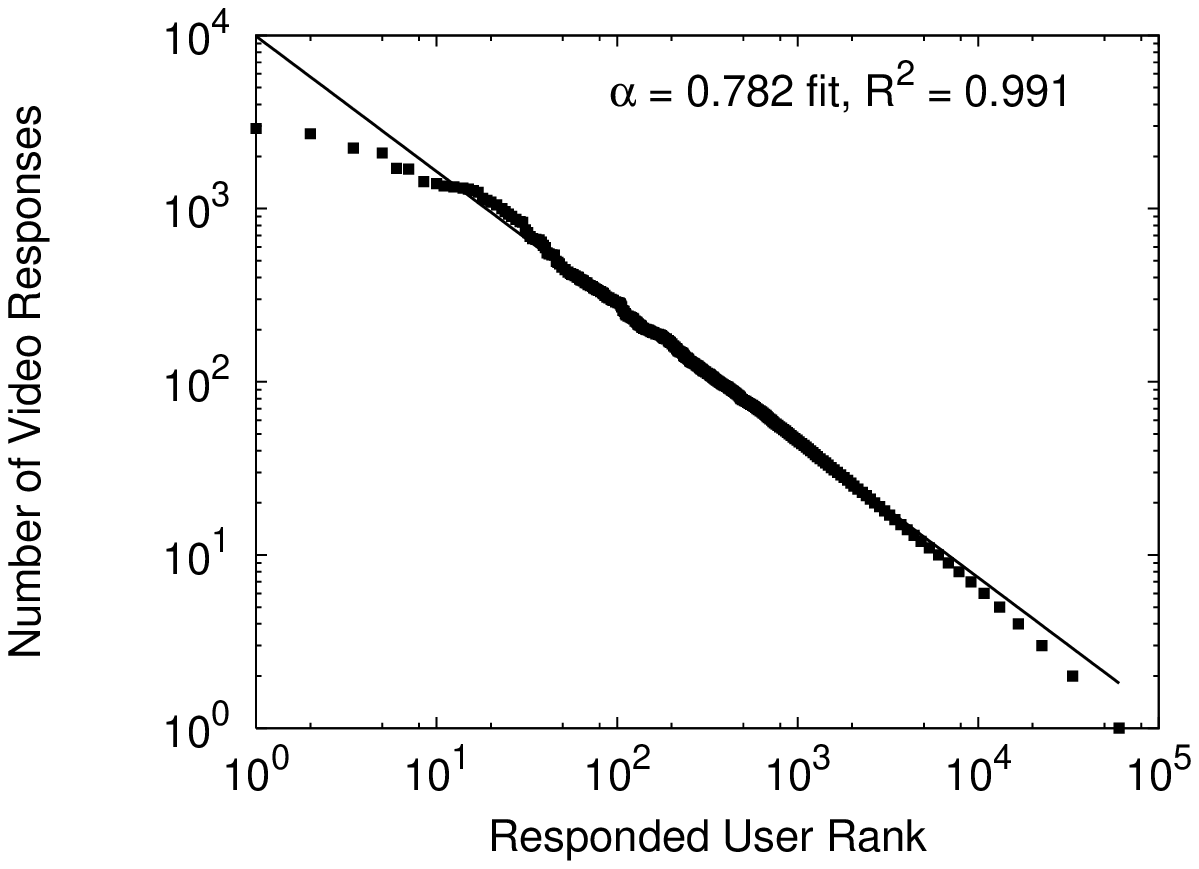}
\includegraphics[width=.27\textwidth]{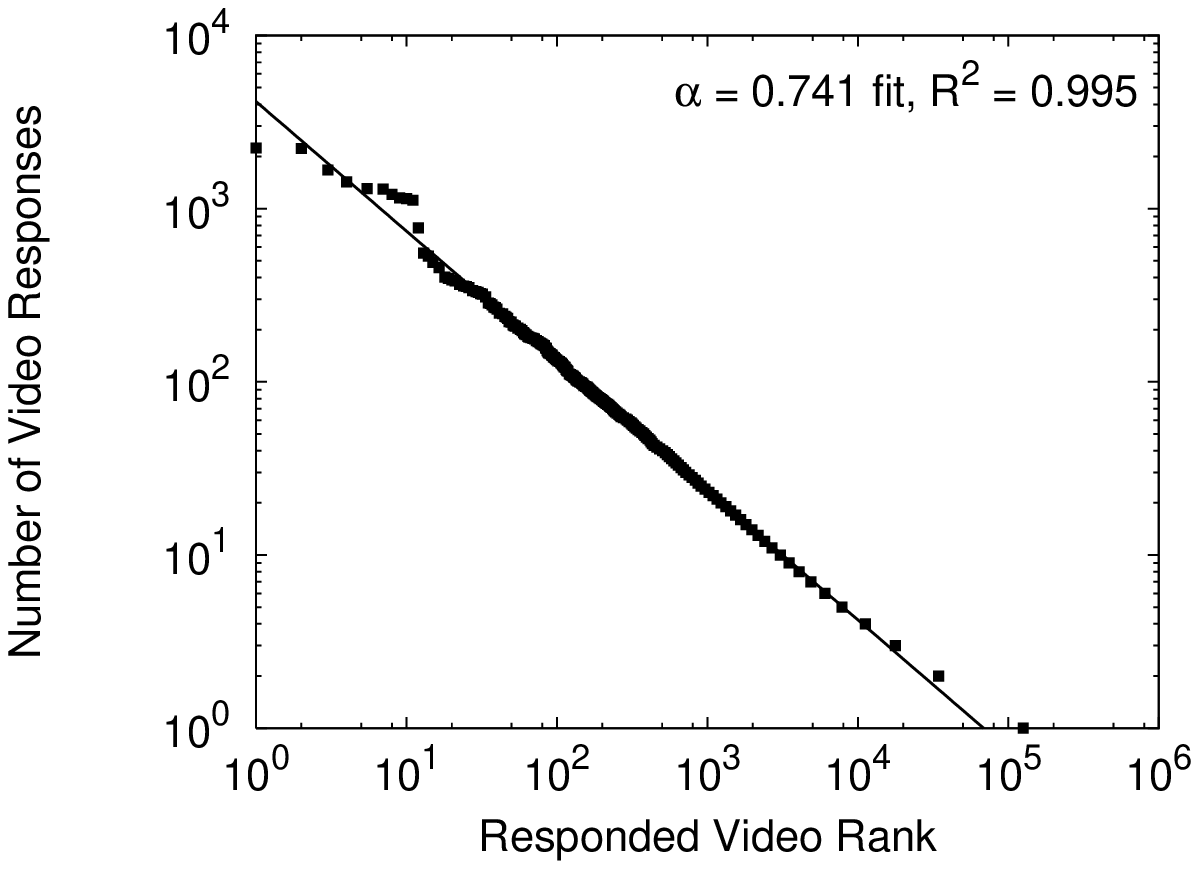}
\caption{Number of Video Responses per Responsive User, Responded User and Responded Video}
\vspace{-0.2cm}
\label{fig:resp_userrank}
\vspace{-0.2cm}
\end{figure*}

\vspace{-0.1cm}
\section{Video Response Characteristcs}
\label{sec:stat}

Figure \ref{fig:resp_userrank} (left) shows the distribution of the number of video responses posted by different users. Note that the 20\% most responsive
users contributed with 65\% of all video responses whereas 84\% of all responsive users posted, each, less than 5 video responses. In fact, the distribution is
well fitted by a power-law distribution (Prob(i posts a video response) $\propto 1/i^{\alpha}$) with $\alpha = 0.568$.  Similarly,
Figures~\ref{fig:resp_userrank} (middle) and (right) show the distributions of the numbers of video responses per responded user and responded video,
respectively. Both distributions are also well fitted by power-laws with $\alpha=0.782$ and $\alpha=0.741$, respectively. Thus, most interactions are initiated by
a few (responsive) users, involve few (responded) users and concentrate around a few videos.

\begin{figure}[t]
\vspace{-0.2cm}
\centering
\includegraphics[width=.3\textwidth]{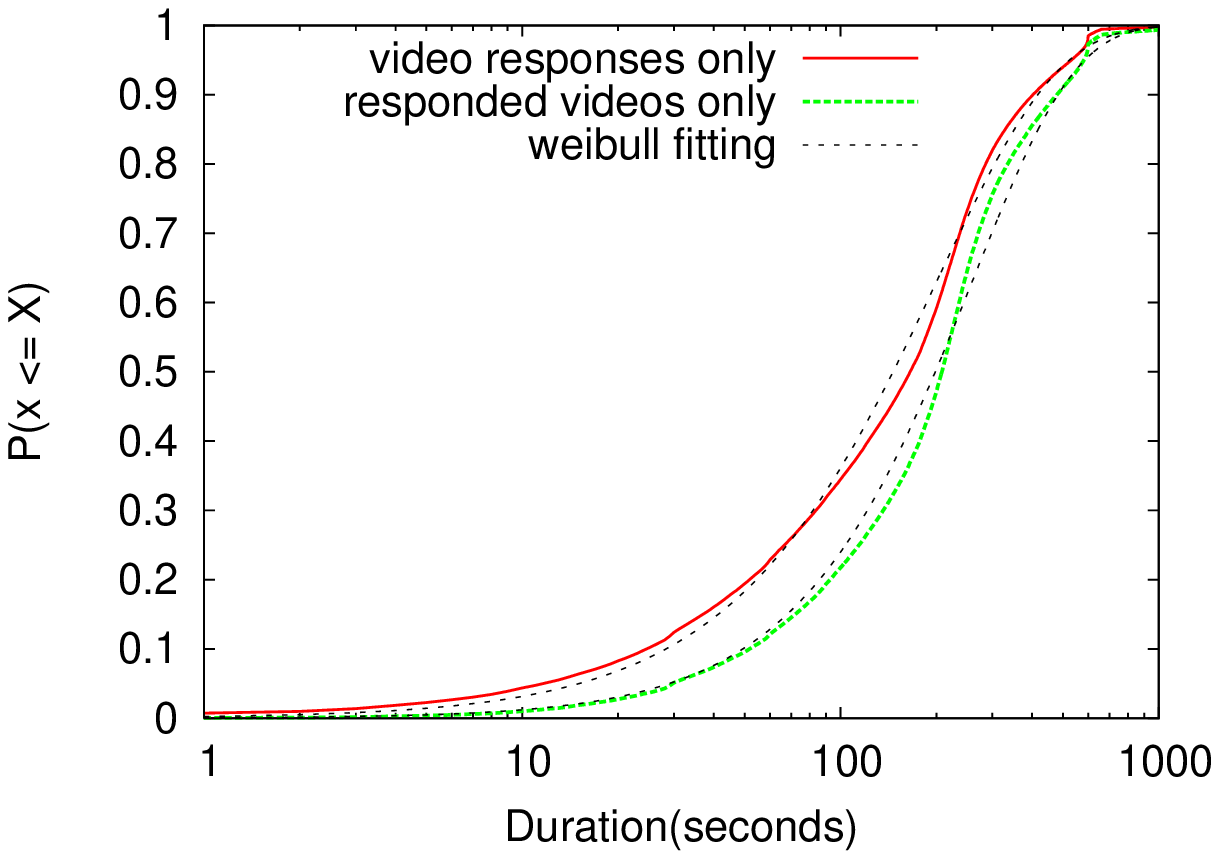}
\vspace{-0.2cm}
\includegraphics[width=.3\textwidth]{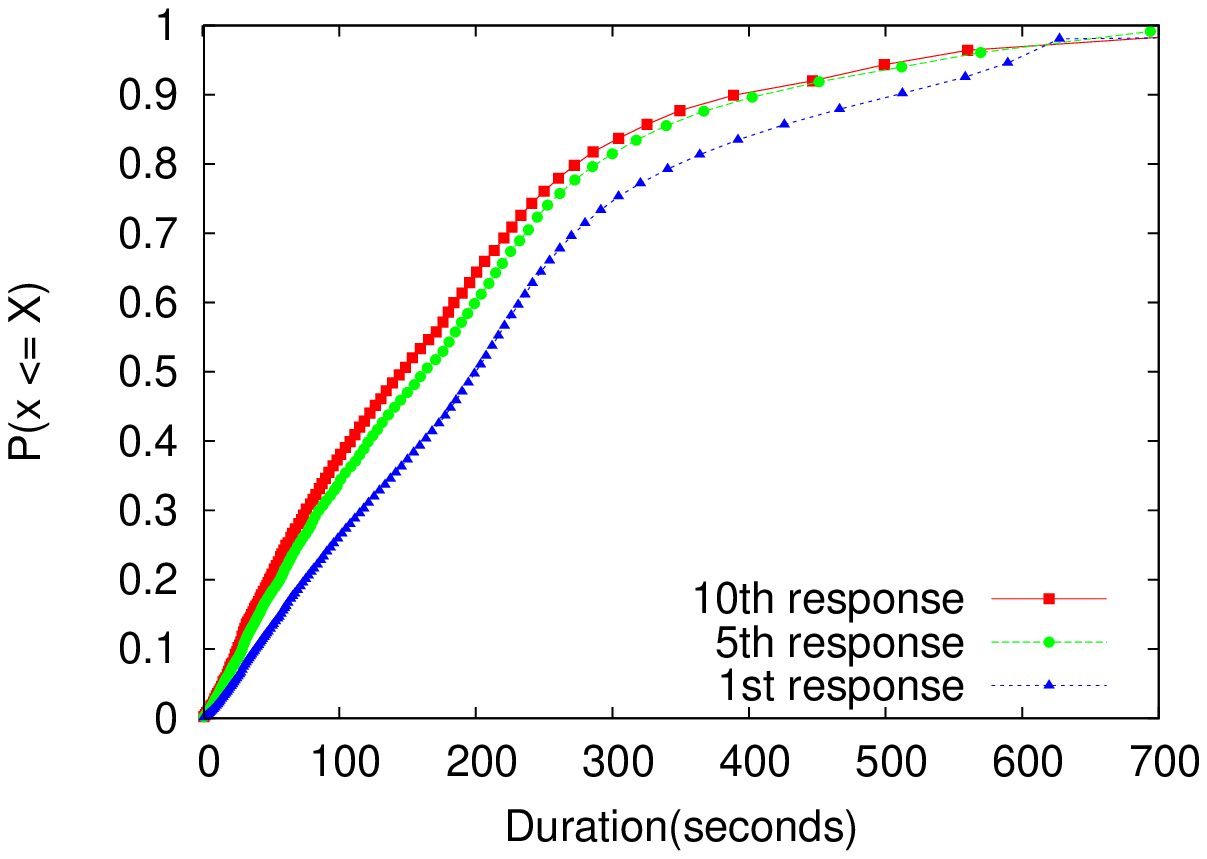}
\caption{Distributions of Duration of Responded Videos and Video Responses}
\vspace{-0.2cm}
\label{fig:duration}
\end{figure}

Figure~\ref{fig:duration} (top) shows the distributions of video durations, considering, separately, only responded videos and only video responses. Both
distributions are very skewed, with 80\% of all samples being under 5 minutes. In fact, both  follow Weibull distributions~\footnote{The pdf for Weibull
distribution is given by: ${f(x) = ({\beta / \alpha})^{-\beta} {(x/ \beta)}^{(\beta-1)} e^{(-x/\alpha)^\beta}}$}, with parameters $\alpha = 0.0023$ and $\beta = 1.15$,
for video responses, and $\alpha = 0.00054$ and $\beta = 1.35$ for responded videos. However, video responses have durations slightly more skewed towards
shorter values. Moreover, we found that, although the duration of a responded video has a small impact on the number of responses it receives (correlation
coefficient $C=-0.008$), there is a strong correlation between the duration of the responded video and the average duration of its responses ($C=0.51$). Longer
responded videos tend to receive longer responses. Moreover, Figure \ref{fig:duration} (bottom) shows that, considering only the $i^{th}$ responses of videos
that had at least $i$ responses, the duration distribution becomes more skewed as $i$ increases.  These results mimic the expected pattern in real-life human
interactions, whereas longer (but interesting) expositions tend to trigger longer replies, initially.  However, replies tend to become shorter as the
interaction progresses, and the discussion dies down.

Interestingly, we found that 25\% of all video responses are self-responses, i.e., responses posted by the user who posted the original video.  Roughly 35\% of
the responded videos received at least one self-response, and around 12\% of them received {\it only} self-responses.  Whereas some of these responses might be
replies to other responses, others might be an attempt at self-promotion, aiming at gaining visibility in order to place video and user in the most-responded
lists. Thus, a question that arises is whether a user can exploit the video response feature to raise its video popularity, i.e., {\it if a video receives many
responses, is it also viewed many times?} The correlation coefficient between the number of responses and the number of views of a responded video shows this is
not often the case ($C = 0.16$). If we disregard all responded videos with at least one self-response, the correlation increases somewhat ($C=0.24$), but
remains low, indicating that the popularity of a video can not be artificially increased by simply adding video responses to it. In other words, posting
(self-)responses aiming at (self-)promotion does not necessarily pay off in YouTube.

A user can post a video response in one of three ways: 1) directly from the user's webcam; 2) choose a video from one of the user's own, pre-existing YouTube
contributions; 3) upload a video from the user's disk drive.  Unfortunately, YouTube does not provide a means to automatically determine in which manner a video
was created. We thus propose to categorize video responses based on the time it was uploaded to YouTube, relative to the upload time  of the  responded video.
We define the Video-Response-Interval (VRI) as the upload time of the response minus the upload time of the responded video.

The cumulative distribution of the VRIs is shown in Figure \ref{fig:nviews} (left).  About 27\% of responses correspond to videos uploaded {\it before} the responded
video, and thus, were certainly not {\it created} as responses to it. This might be explained by the video content itself. For instance, we observed that
one of the responded  videos having many previously uploaded responses explicitly actually requested existing YouTube videos for responses. On the other hand,
it might also be the result of users uploading existing (and not necessarily related) videos as self-responses.  Moreover, 42\% of responses are added within the
first month after the responded video was uploaded, indicating a prompt reaction from responsive users. Nevertheless, a non-negligible fraction (17\%) of
responses are added long after the responded video appeared in the system (i.e., VRI $\geq$ 100 days), meaning that some videos exhibit long-term popularity,
and new interactions are initiated long after it was uploaded.

\subsection{Video Response Popularity Distribution}

\begin{figure}[t]
\vspace{-0.2cm}
\centering
\includegraphics[width=.23\textwidth]{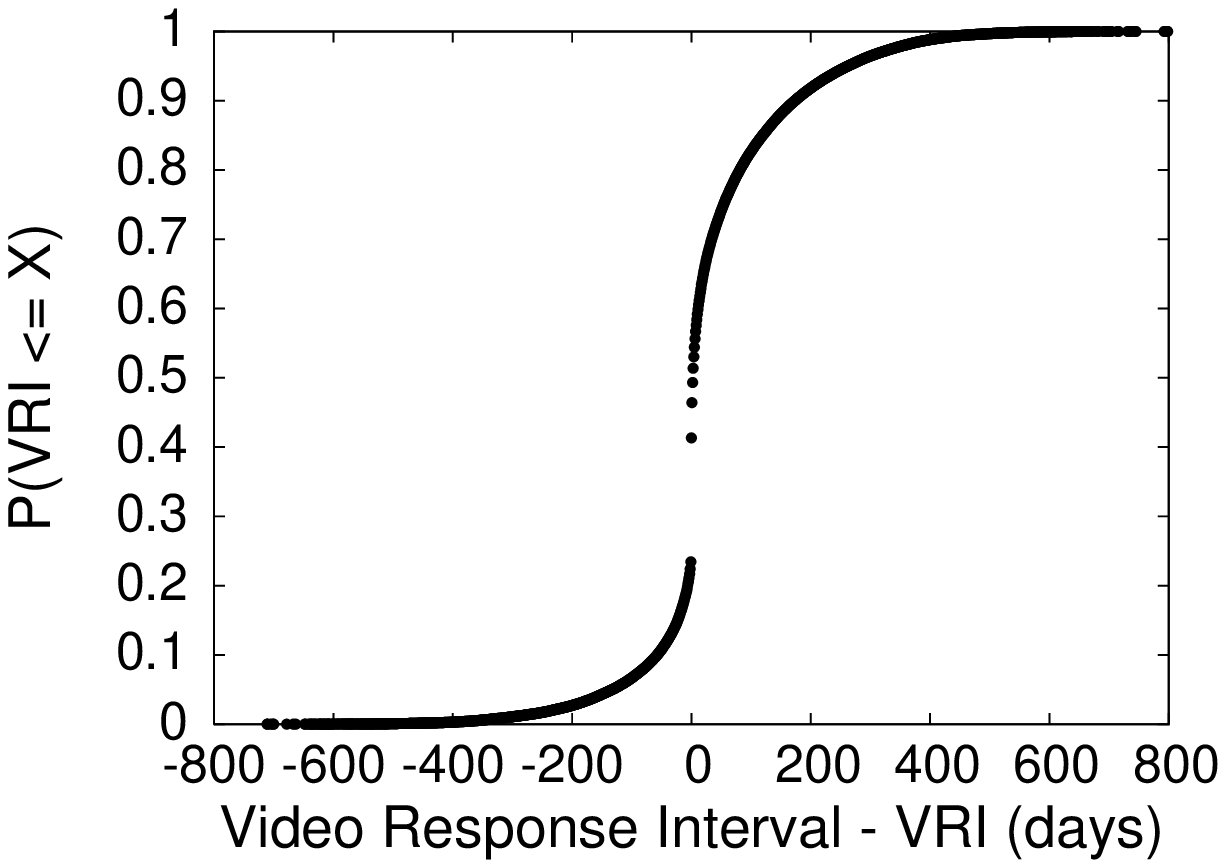}
\includegraphics[width=.23\textwidth]{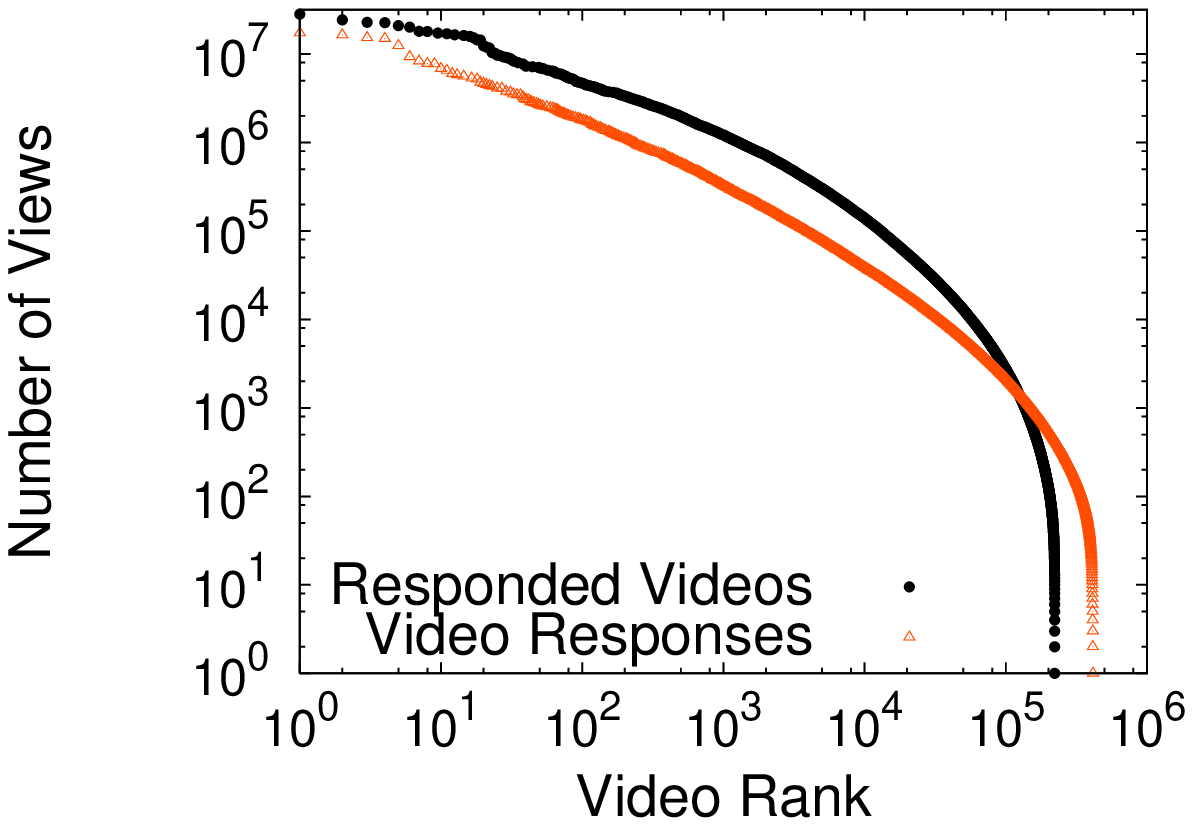}
\caption{Distribution of VRIs and Rank Distribution of the Number of Views for Responded Videos and Video Responses}
\vspace{-0.2cm}
\label{fig:nviews}
\end{figure}

Figure~\ref{fig:nviews} (right) shows the rank distribution of responded videos and video responses in terms of number of views in YouTube.  We observe that the total
numbers of views for responsive and responded videos are not dominated by a very small number of very popular videos. From Figure.~\ref{fig:nviews} (right), we calculated
that there are more than 33,000 video responses and 57,000 responded videos that have been viewed more than 10,000 times each. The median number of views for
the video responses and responded videos is 508 and 2,028 respectively, while 90\% of video responses and responded videos we crawled were viewed more than 61
and 129 times, respectively. This in turn raises interesting questions about the consumption behavior of YouTube. Note that the number of views reported include
both complete and incomplete views (where users stopped viewing after a few seconds or more). We conjecture that due to the low effort and cost of viewing a
video and the rich interconnections between videos on the site, there is a large amount of fairly random surfing and exploration where visitors and users
serendipitously explore videos. This could explain the large number of views of even unpopular videos and contrasts with traditional Web content, dominated by
Zipf's behavior.  One may conjecture  that a video service that charges even a small amount per video started would exhibit a very different behavior.

\subsection{Video Response Geographical Distribution}

\begin{figure}[t]
\vspace{-0.2cm}
\centering
\includegraphics[width=.23\textwidth]{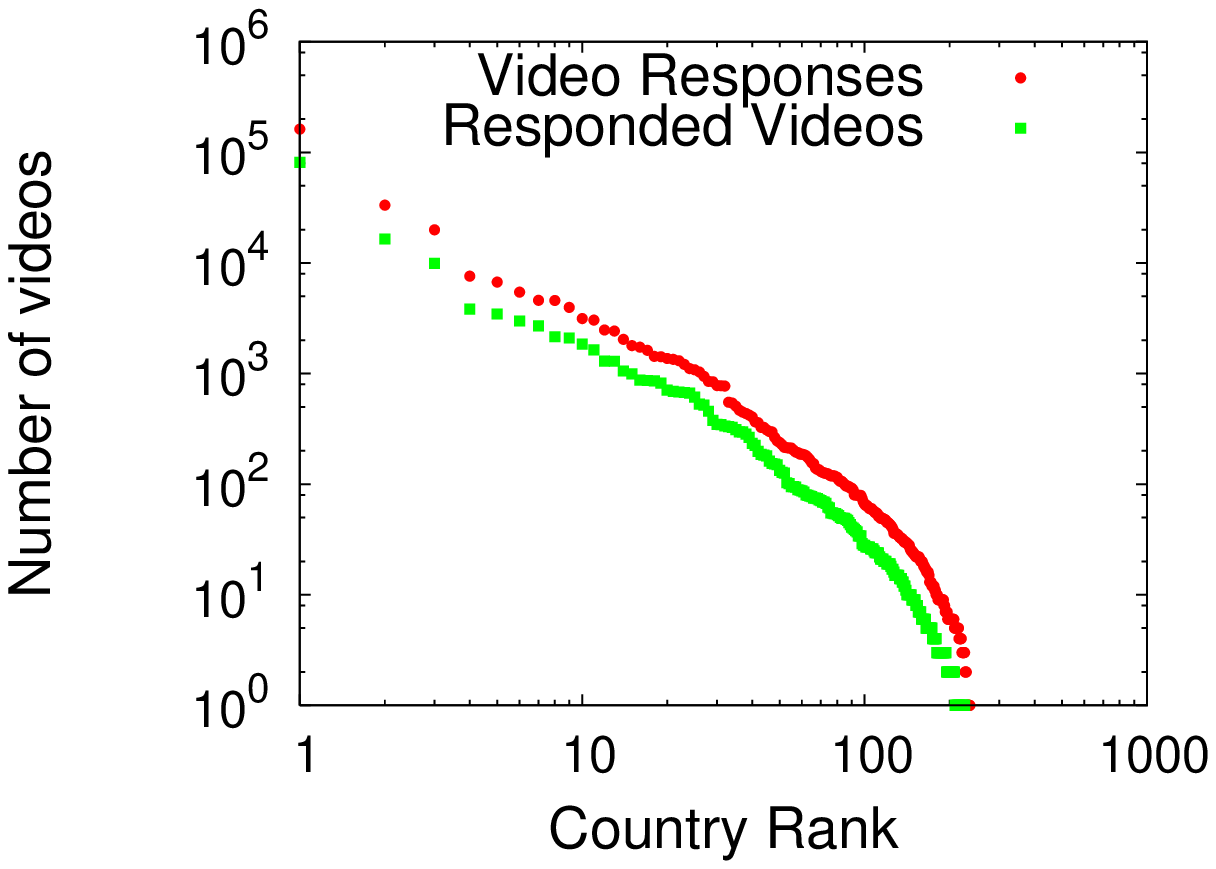}
\includegraphics[width=.23\textwidth]{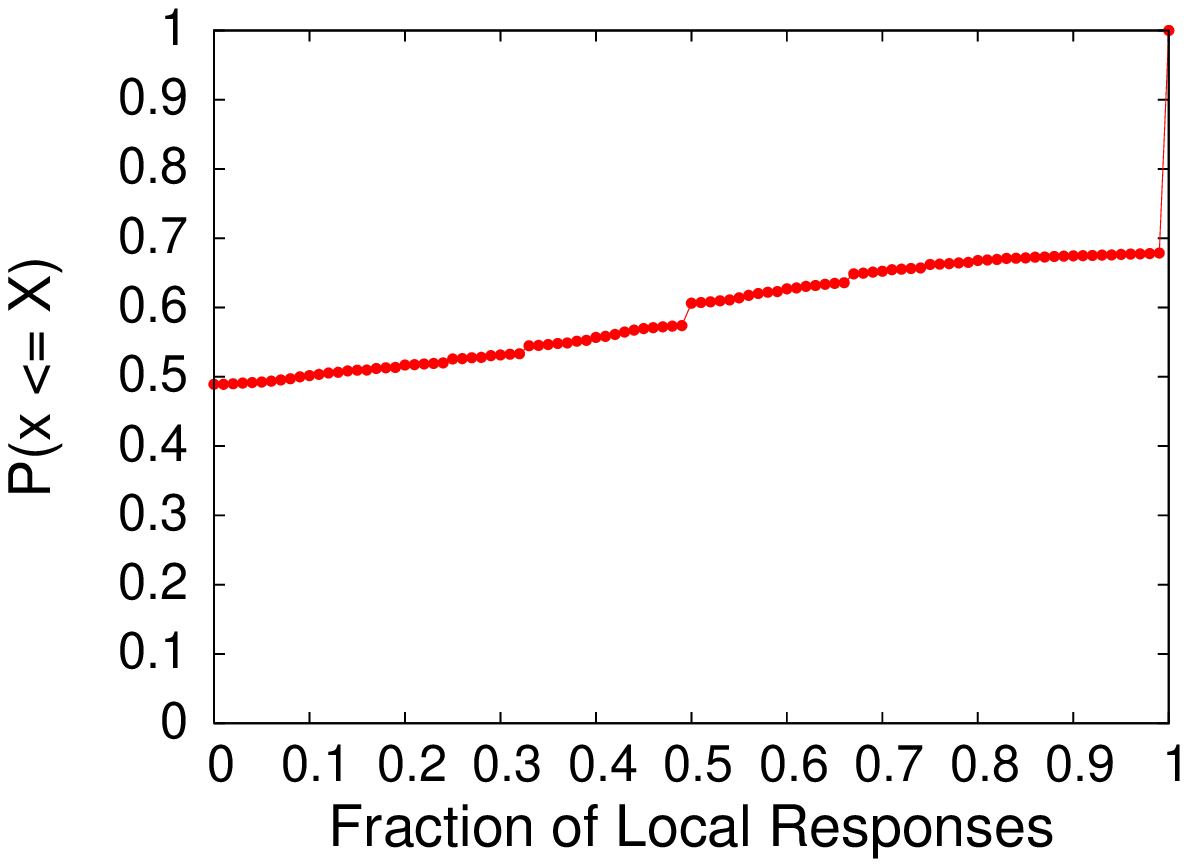}
\vspace{-0.2cm}
\caption{Distribution of Video Responses over Countries}
\vspace{-0.2cm}
\label{fig:countries}
\end{figure}

An important question that is often asked regarding web characterization studies has to do with the geographical representativeness of the sample. Geography and
friendship have been used to build real social network models~\cite{liben2005}. As evident from Figure~\ref{fig:countries} (left), the sample we characterize 
is fairly large in terms of the number of countries (as identified by the country described in the user profile) and the number of video responses
and responded videos uploaded by users of different countries. Using the country identification, we are able to map the contributor population to over 236
countries. The top five countries in our sample account for 76.8\% of total video responses uploaded to YouTube. The plots suggest a power law-like profile,
with parameter $\alpha = 2.12$ ($R^2 = 0.92$) and  $\alpha = 2.22$ ($R^2 = 0.93$) for video responses and responded videos, respectively. We also defined the
percentage of local video responses as the ratio of video responses from the same country of the responded video to the total number of video responses.
Figure~\ref{fig:countries} (right) shows the cumulative distribution of the percentage of local video responses over all responded videos of our sample.  We
notice that more than half of the ``video conversations'' involves at least some participants from the same country of the original contributor.  For example,
40\% of responded video has a percentage of local responses superior to 60\%, which may be useful information for designing content distribution mechanisms.

\subsection{Video-Based Interactions}

Our characterization of user interactions at the video level consists of examining the sequence of video responses following a YouTube video. For each responded
video $V_i$, let $n_i$ denote the number of video responses, and denote the sequence of responses as $\{VR_{i,1}; VR_{i,2}; ...; VR_{i,j}; ... ; VR_{i,n_i} \}$,
where $VR_{i,j}$ is the $j^{th}$ video response.

A user may add multiple video responses in sequence to the same video. We define a sequence of responses $S_{i,k}$ as a series of {\em consecutive} responses
from the same user to video $V_i$. Thus, the ordered list of video responses to video $i$ can also be expressed as $\{S_{i,1}; S_{i,2}; ... S_{i,s_i} \}$, $1
\le s_i \le n_i$, where $S_{i,k}$ is the sequence $\{ VR_{i,j}; VR_{i,j+1} ... \}$ of {\it consecutive} responses from user $U_{i,k}$.  Note that the same user
may post multiple (non-consecutive) sequences of responses  to the same video $V_i$, i.e., we may  have $U_{i,k} = U_{i,j}$ for $k$ $\not\in$ [j-1,j+1].

In the following example, video $V_i$ received 7 video responses grouped into 4 sequences, posted by 3
users ($U_{1}$, $U_{2}$ and $U_{3}$):
$$
V_i; \overbrace{\underbrace{VR_{i,1}, VR_{i,2}}_{U_1}}^{S_{i,1}},\overbrace{\underbrace{VR_{i,3}}_{U_2}}^{S_{i,2}}, \overbrace{\underbrace{VR_{i,4}, VR_{i,5},VR_{i,6}}_{U_1}}^{S_{i,3}}, \overbrace{\underbrace{VR_{i,7}}_{U_3}}^{S_{i,4}}
$$
Our characterization focuses on a simple metric,
defined as the ratio U/S of the {\em Number of Unique Responsive Users} to the
{\it Number of Sequences of Responses}.  In the above example, this ratio is 0.75.  A video-based interaction with a ratio U/S close to 0 indicates an
asynchronous video dialogue between a relatively small number of highly active users, who keep the discussion alive with multiple (not necessarily consecutive)
responses to each other.  This type of interaction is akin to the exchanges and debates in a parlor or public forum, in which the communication underscores a
many-to-many dialogue between participants.  At the other extreme, when the ratio  approaches one, there may be two types of interaction.  One type occurs when
the number of unique responsive users  equals the number of sequences of responses, resembling a register, petition or guest-book, for which the communication
is many-to-one, and the purpose of a video response is to record a comment (or support a petition, etc.). The other type has just one user posting a single
sequence.

\begin{figure}[t]
\vspace{-0.2cm}
\centering
\includegraphics[width=.3\textwidth]{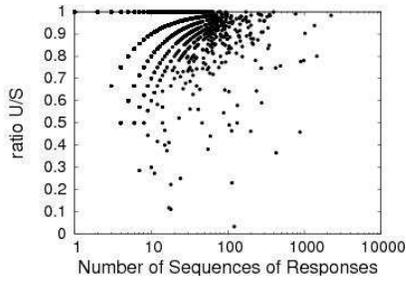}
\caption{Types of video-based interactions}
\vspace{-0.2cm}
\label{fig:pvi}
\end{figure}

Figure~\ref{fig:pvi} shows the ratio U/S versus the number of sequences of responses for each responded video in our data set.  Different types of
interactions can be seen across all responded videos, characterized by a wide range of U/S values. However, only a relatively small number of videos triggered
lively interactions, with each responsive user adding on average at least 3 sequences of responses.  In fact,  the vast majority of the responded videos (99\%)
triggered interactions with only  one sequence of responses per responsive user  (i.e., ratio equal to 1).  In fact, Figure \ref{fig:pvi} shows these
interactions occur among groups of responsive users of  varying sizes (i.e., number of sequences). 

\vspace{-0.1cm}
\section{Network Characteristics}
\label{sec:social}

Out of the several networks that emerge from the user interactions enabled by YouTube features,  we select the user graph $(A,B)$ (see Section 2) for an
in-depth analysis.  Table~\ref{tab:workload} presents the main statistics of the  network built from the {all-time top-100}  data set and for its largest
strongest connected component.

\begin{table}[b]
\vspace{-0.2cm}
\begin{center}
\scriptsize
\begin{tabular}{|r|l|l|l|l|}
\hline
\bf Characteristic & \bf Dataset & \bf Largest SCC\\
\hline
\# nodes &  160,074 &   7,776 \\
\# edges & 244,040 &  33,682 \\
Avg Clustering Coefficient & 0.047 & 0.137 \\
\# nodes of largest SCC  & 7,776 &   7,776 \\
\# components    & 149,779 & 1 \\
$r $ & -0.017 &   0.017 \\
Avg distance &  8.40  &   8.40 \\
Avg $k_{in}$ (CV) &  1.53 (9.38)    &  4.33 (3.14) \\
Avg $k_{out}$ (CV) &  1.53 (1.717)   &    4.33 (1.28)  \\
Avg $k$  &  3.06   &  8.66 \\
\hline
\end{tabular}
\end{center}
\vspace{-0.2cm}
\caption{Summary of the Network Metrics}
\vspace{-0.2cm}
\label{tab:workload}
\end{table}

\subsection{Degree Distribution}

\begin{figure}[b]
\vspace{-0.2cm}
\centering
\includegraphics[width=.23\textwidth]{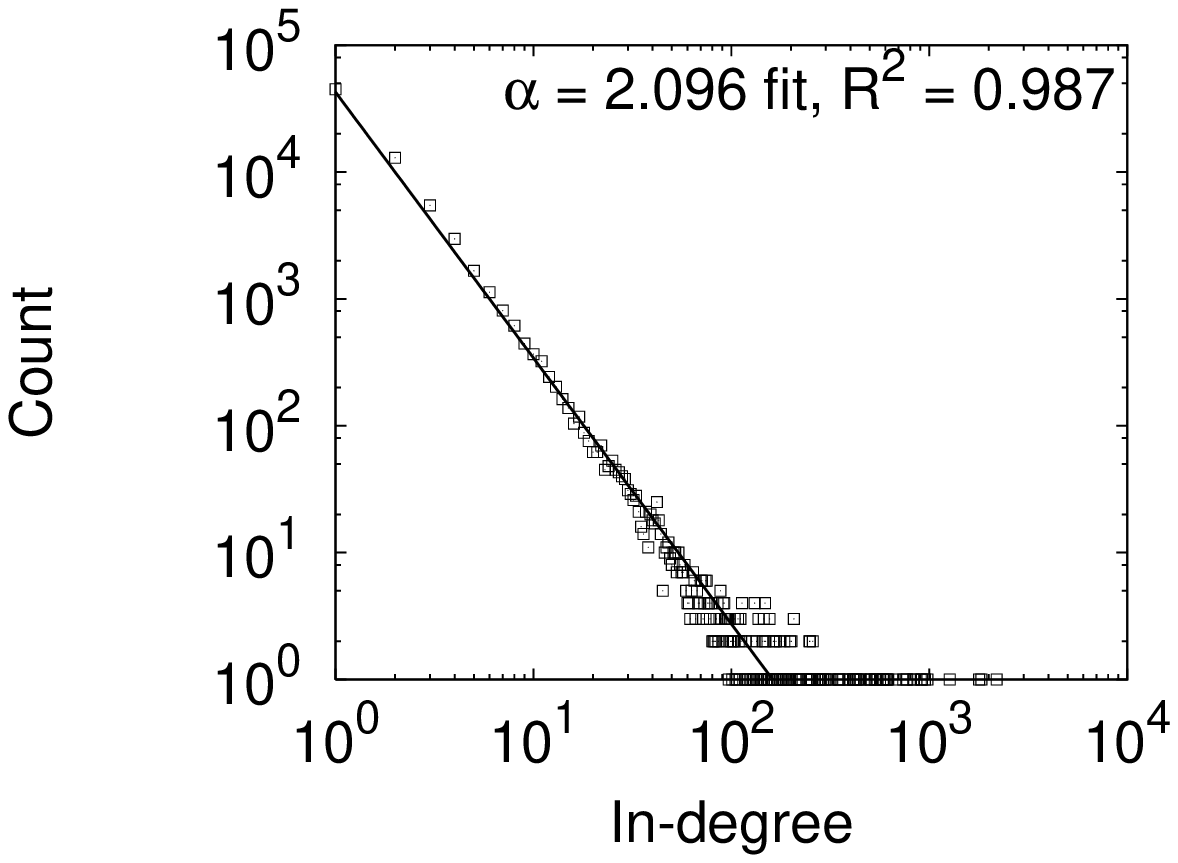}
\includegraphics[width=.23\textwidth]{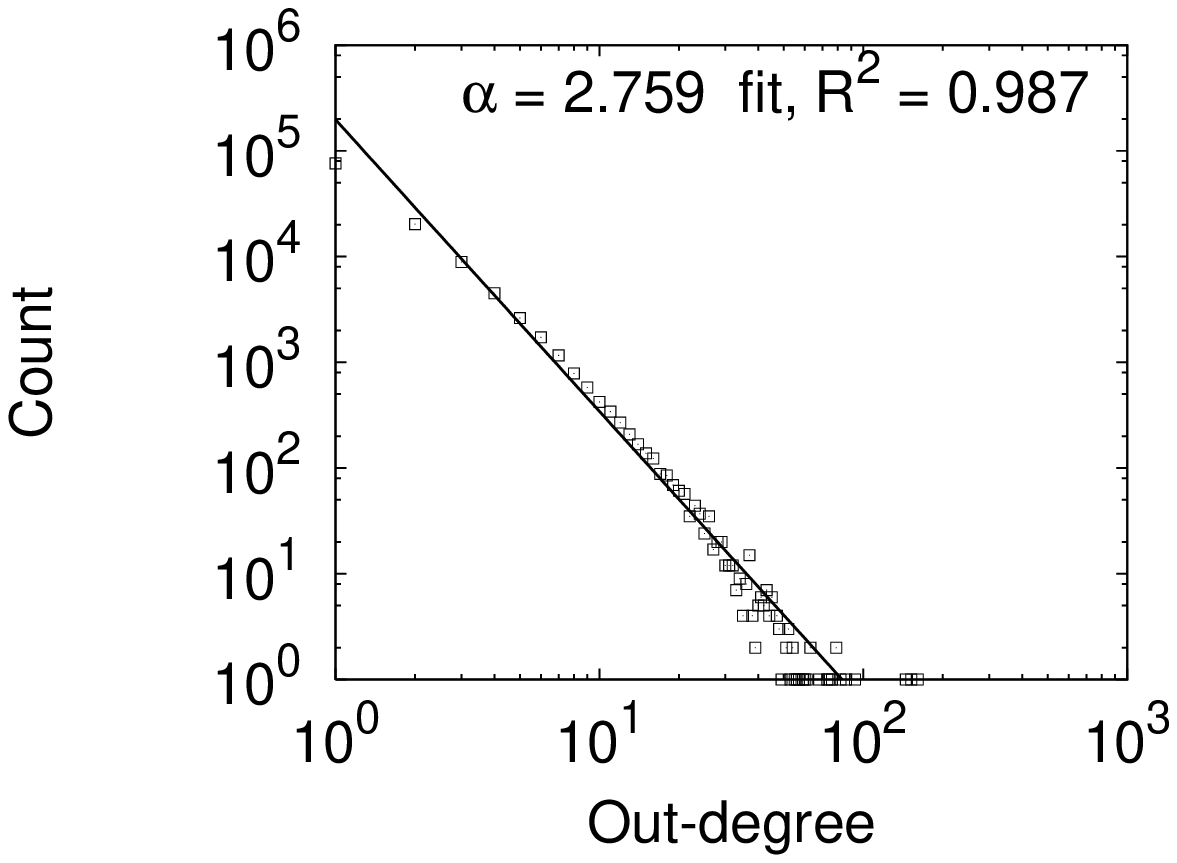}
\caption{In-Degree and Out-Degree Distributions.}
\vspace{-0.2cm}
\label{fig:inoutdegree}
\vspace{-0.2cm}
\end{figure}

The key characteristics of the structure of a directed network are the in-degree $(k_{in})$ and the out-degree $(k_{out}$) distributions. As shown in
Figure~\ref{fig:inoutdegree}, the distributions of the degrees for the entire graph follow power laws $P(k_{in, out})\propto 1/k_{in, out}^{\alpha^{in, out}}$,
with exponent $\alpha^{in}=2.096$ and  $\alpha^{out}=2.759$  with the following coefficient of determination: $R^2 = 0.98$ and $R^2 = 0.97$.   The scaling
exponents of the whole network lie in a range of 2.0 and 3.4, which is a very common range for social and communication networks~\cite{ebel}. Our results agree
with previous measurements of  many real-world networks that exhibit power law distributions, including the Web and social networks defined by blog-to-blog
links~\cite{tomkins2003}.

\begin{figure}[b]
\vspace{-0.2cm}
\centering
\includegraphics[width=.23\textwidth]{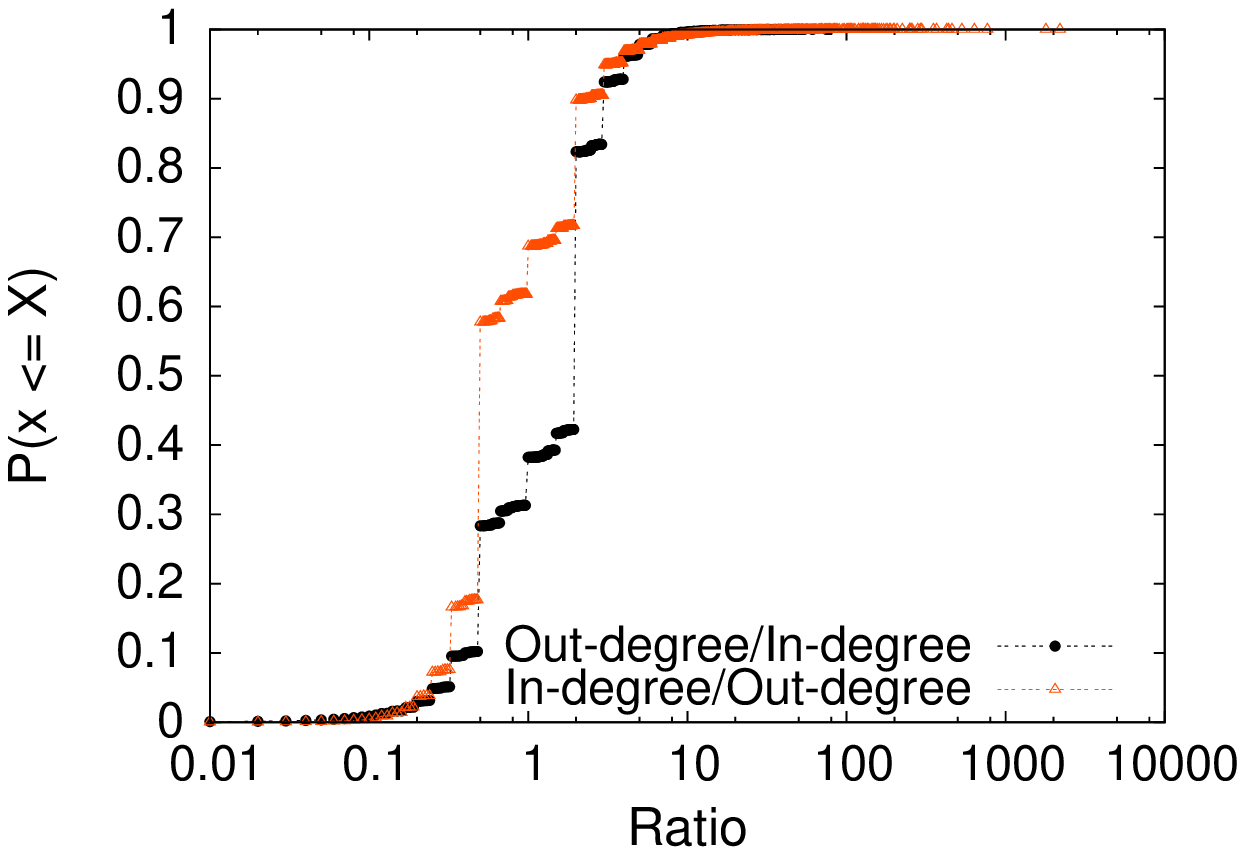}
\includegraphics[width=.23\textwidth]{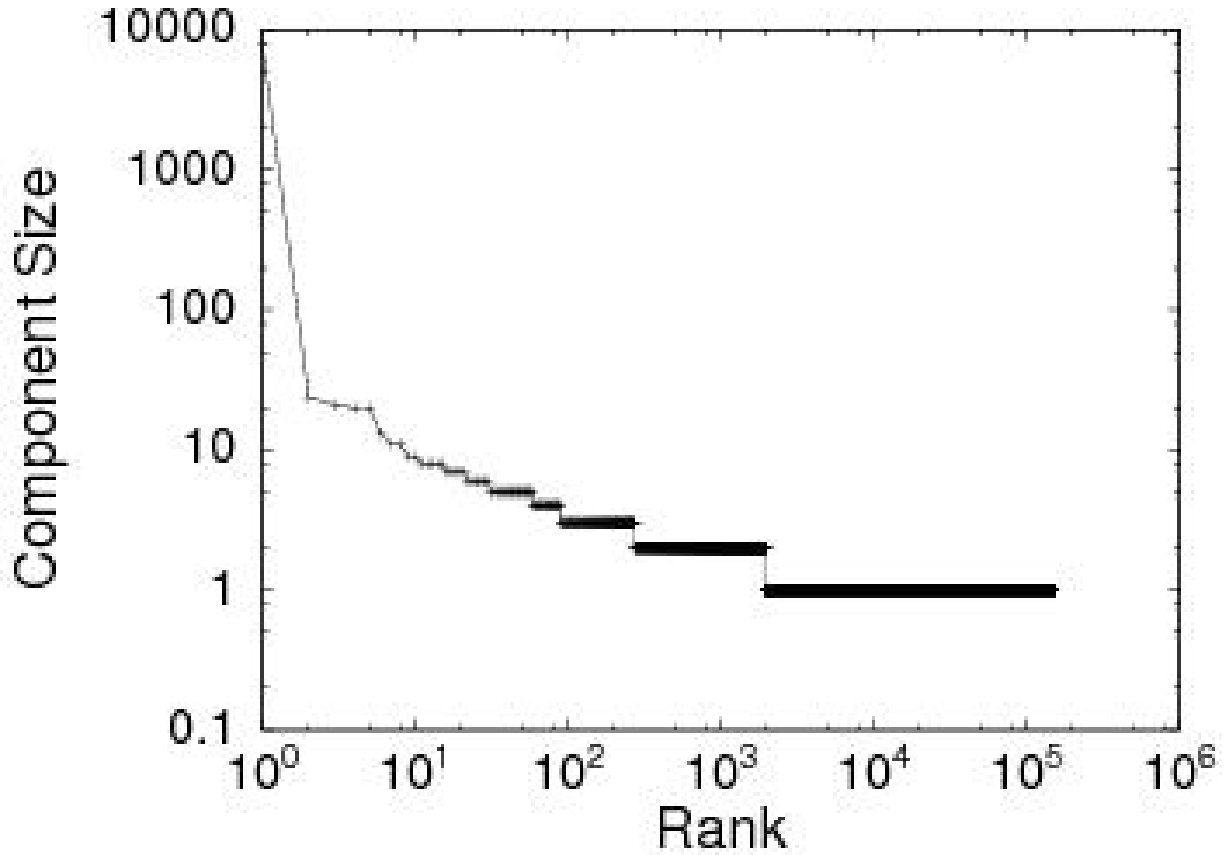}
\caption{CDF of in-degree to out-degree ratio (left). Component Size vs Rank (right).}
\vspace{-0.2cm}
\label{fig:inoutratio}
\end{figure}

The in-degree exponent is smaller than  the exponent of the out-degree distribution, indicating that there are more users with larger in-degree than out-degree.
This fact suggests a link asymmetry in the directed interaction network. Unlike other social networks that exhibit a significant degree of
symmetry~\cite{mislove2007}, the user interaction network shows a structure similar to the Web graph, where pages with high in-degree tend to be authorities and
pages with high out-degree  act as hubs directing users to recommended users~\cite{kleinberg1999}.  In order to investigate this point further,
Figure~\ref{fig:inoutratio} (left) shows the cumulative distribution of ratios between in-degree and out-degree  for the user interaction network. The network has 60\%
of the users with out-degree higher than in-degree and 5\% of the users with significantly higher in-degree than out-degree.  This is evidence that a few users
act as ``authorities'' and ``hubs''. Interestingly, we have observed in our dataset that authority-like users (that is, 
highly responded users), with high
in-degree, are typically media companies that upload professional content, including sports, entertainment video and TV series.  
This type of node in our
network receives video response from many  YouTube users.  Nodes with  very high out-degree  may indicate either very active users or spammers, i.e., users that distribute  content  (i.e., video responses) that "legitimate" users have not solicited.

According to~\cite{Newman2003}, assortative mixing and high clustering coefficient are two graph theoretical quantities typical of social networks. We now
investigate assortative mixing. Clustering Coefficient is analyzed in the next section. A network is said to exhibit assortative mixing if the nodes with many
connections tend to be connected to other nodes with many connections. Social networks usually show assortative mixing. The assortative (or disassortative)
mixing is evaluated by the Pearson coefficient $r$, which is calculated as follows.~\cite{Newman2002b}:
\begin{eqnarray}
\label{eq:pearsoncoefficient}
r = \frac{ \sum_i{j_i k_i}  -  M^{-1} \sum_i{j_i} \sum_{i'}{k_{i'}}} {\sqrt{[\sum_i{j_{i}^{2}} - M^{-1}(\sum_i{j_ i})^2][\sum_i{k_{i}^{2}}  - M^{-1} (\sum_i{k_i})^2]}},
\end{eqnarray}
where $j_i$ and $k_i$ are the excess in-degree and out-degree of the vertices that the {\it i}th edge leads into and out of,
respectively, and $M$ is the total number of edges in the graph.

Table~\ref{tab:workload} shows values of $r$ for the directed graph of the interaction network. The  video-response network has a disassortative mixing $r=-0.017$,
where high degree nodes preferentially connect with low degree ones and vice versa. By contrast social networks have significant assortative mixing, which
accords with the notion of social communities.  So, the entire user interaction graph does not show evidence of the formation of a large social community.

\subsection{Clustering Coefficient}

\begin{figure}[t]
\vspace{-0.2cm}
\centering
\includegraphics[width=.23\textwidth]{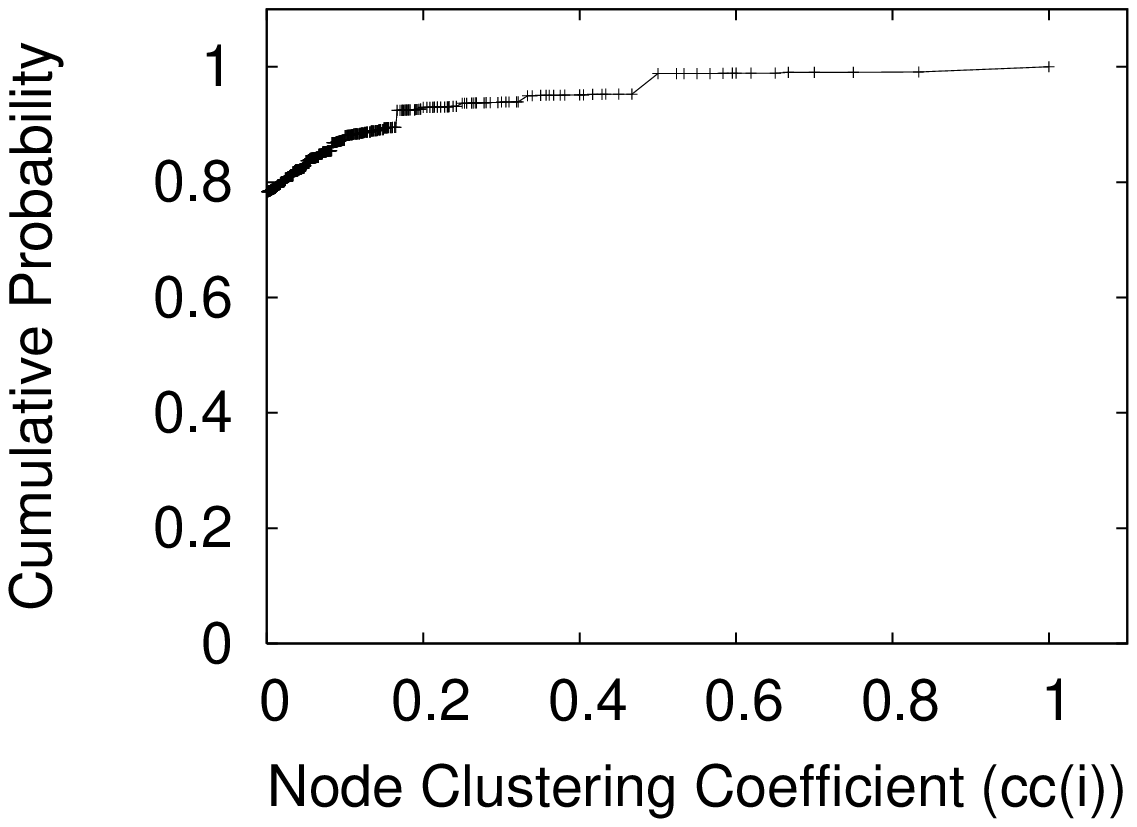}
\includegraphics[width=.23\textwidth]{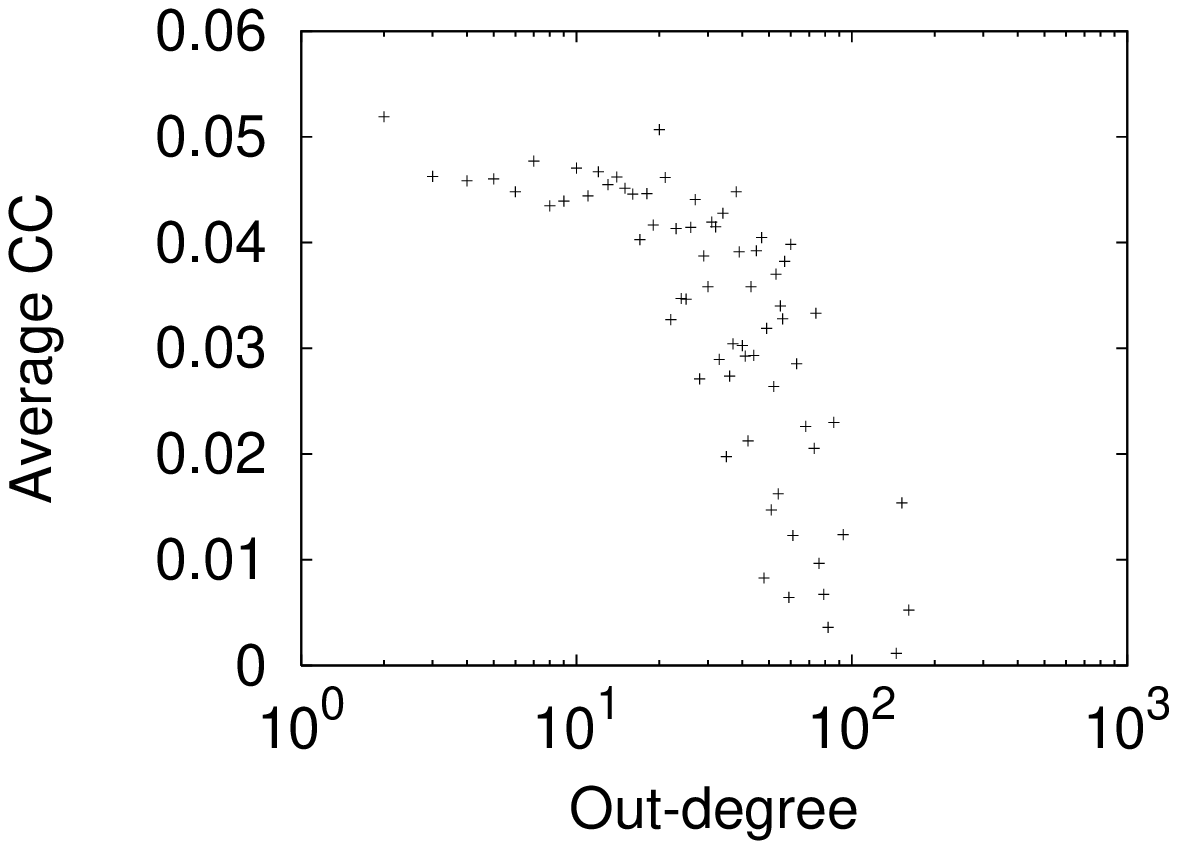}
\caption{Distribution of $CC$ (left) and avg. $CC$ node out-degree (right).}
\vspace{-0.2cm}
\label{fig:ccperdegree}
\vspace{-0.2cm}
\end{figure}

It has been suggested in the literature that social networks possess a topological structure where nodes are organized into communities~\cite{Newman2003}, a
feature that can account for   the  values for the clustering coefficient and degree correlations.  The clustering coefficient of a node $i$, $cc(i)$ is the
ratio of the number of existing edges over the number of all possible edges between $i$'s neighbors.  The clustering coefficient of a network, $CC$, is the mean
clustering coefficient of all nodes. The average CC over the whole network is $CC=0.047$, whereas the mean clustering coefficient for a random graph with identical degree distribution but random links is $CC=0.007$, which shows the presence of small communities in the video-response network.  The leftmost part of Figure~\ref{fig:ccperdegree} shows the
cumulative distribution of the clustering coefficient.  The network contains a significant fraction of their nodes with zero clustering coefficient.
Specifically, $80\%$ of all nodes in the entire user interaction network have $CC=0$.  This feature  indicates that there is a clear difference on average
between clustering in the entire network and the components of the network. The right part of the figure shows how the clustering coefficient varies with the
node out-degree. Higher values of the clustering coefficient occur among low degree-nodes, suggesting the lack of large communities around high-degree nodes.
Our conjecture is that highly responsive users do not necessarily have social links with the contributors of the videos that they are responding to.  Therefore,
there may not exist a sense of community among  the users that receive video responses from a single responsive user.
Low degree nodes might explain the formation of very small communities, composed of a few people like a family or a group of friends that share videos and interact through video responses.

Table~\ref{tab:workload} shows the size of the largest strongly connected component (SCC) and the total number of other strongly connected components of the video response user graph ($A$, $B$).
Figure~\ref{fig:inoutratio} (right) shows the distribution of the size of the networked components sorted  from the largest component to the smallest one. The distribution suggests a general structure that includes the largest SCC, the middle components (i.e., 1974), and a large number of components with just one node (i.e., 147805).
As we are working with a directed graph, these components with size one are nodes with links in only one direction.
The middle components are tightly connected groups of users, representing  small size communities (e.g., families and groups of friends) that express their
interests and establish communication via video responses. The largest SCC represents about 5\% of the nodes, but it is
considerably larger than the others. It concentrates 10\% of the views and 22\% of the video responses and deserves further analysis.  Although it includes
about 5\% of the nodes, its size is comparable to the size of SCC in other networks, derived from blogosphere
samples~\cite{adamic07b}. The differences in size of connected components may be due to time factors, that account for the adoption by users  of specific
features (i.e., video response) in social networking environments.  In order to understand its characteristics, we investigate  network  properties of the giant
component in the video response user graph. The average clustering coefficient of the giant component  is $CC=0.137$, three times greater than the clustering
coefficient of the entire network. Thus, user interactions captured by the giant component might form a more tightly connected community.

\vspace{-0.1cm}
\section{Conclusion}
\label{sec:conc}

We crawled YouTube to obtain a large representative subset of the video response user graph. Our characterization was done from two perspectives:
video response view and interaction network view.  In addition to providing statistical models for various characteristics (popularity profiles, duration,
geographical, etc.), our study has unveiled a number of interesting findings. For example, the characteristics of social video sharing services are
significantly different from those of traditional stored object workloads, based on text and image. Part of the difference stems from the change from textual
communication to stream-based communication, creating a new paradigm for online communication.  

Our current and future work is focused on leveraging many of the findings and conclusions presented in this paper along a number of dimensions. First, we are
looking into using the characteristics found in this paper as models to be used for designing mechanisms for content distribution (i.e., pre-fetching,
replication and caching) based on social network characteristics.  For instance, if a particular stream is posted by a user, then for a distribution server at a
particular location, we can associate that stream with the social community of that individual, identified by the video-response network.  Second, we are evaluating
the use of network node characteristics (i.e., degree distribution, degree correlation, etc)  to identify spam in online social networks~\cite{molina2007}.

\bibliographystyle{abbrv} \bibliography{wosn2008.bib}


\end{document}